\documentclass[12pt,preprint]{aastex}

\begin{document}

\title{Discovery of an M4 Spectroscopic Binary in Upper Scorpius:
A Calibration Point for Young Low-Mass Evolutionary Models}


\author{Ansgar Reiners\altaffilmark{1} \and Gibor Basri}
\affil{Astronomy Department, University of California, Berkeley, CA 94720
\email{[areiners, basri]@astron.berkeley.edu}}
\altaffiltext{1}{Hamburger Sternwarte, Universit\"at Hamburg, Gojenbergsweg 112, D-21029 Hamburg, Germany}

\and

\author{Subhanjoy Mohanty}
\affil{Harvard-Smithsonian Center for Astrophysics, Cambridge, MA 02138} 
\email{smohanty@cfa.harvard.edu}


\begin{abstract}
  We report the discovery of a new low-mass spectroscopic (SB2)
  stellar binary system in the star-forming region of Upper Scorpius.
  This object, UScoCTIO~5, was discovered by \cite{Ardila00}, who
  assigned it a spectral class of M4.  A Keck~I HIRES spectrum
  revealed it to be double-lined, and we then carried out a program at
  several observatories to determine its orbit.  The orbital period is
  34 days, and the eccentricity is nearly 0.3.  The importance of such
  a discovery is that it can be used to help calibrate evolutionary
  models at low masses and young ages.  This is one of the outstanding
  problems in the study of formation mechanisms and initial mass
  functions at low masses.  The orbit allows us to place a lower limit
  of $0.64 \pm 0.02$\,M$_\odot$ on the total system mass. The
  components appear to be of almost equal mass.  We are able to show
  that this mass is significantly higher than predicted by
  evolutionary models for an object of this luminosity and age, in
  agreement with other recent results.  More precise determination of
  the temperature and surface gravity of the components would be
  helpful in further solidifying this conclusion.
\end{abstract}
\newpage
\section{Introduction}

The study of very low mass stars and brown dwarfs has experienced
rapid growth during the past 15 years, primarily due to the advent of
the next generation of large telescopes. Many questions are beginning
to be answered about the atmospheric and evolutionary state of these
objects, yet many fundamental questions remain. Thus far we have not
gathered much empirical data on mass, one of the most fundamental
quantities of interest. The evolutionary models that are currently
used to assign almost all masses, based on age and luminosity, are
largely untested. Modelers warn that such models cannot be trusted too far
\citep{Baraffe03} for very low masses and very young ages. The search
for binary systems that would provide fundamental stellar parameters
is only now beginning to bear fruit. Indeed, the frequency of
occurrence and physical configurations of very low mass binaries are
themselves not well understood. These systems appear to have a
somewhat, but not drastically, lower binary frequency than solar-type
stars, and the scale of the systems seems to be smaller, along with
more equal mass ratios \citep{Bouy03}.

Recently there have been indications that the evolutionary models
substantially under-predict masses for young objects in the 0.03-0.5
M$_\odot$ range. \cite{Hillenbrand04} compiled a collection of 115
low-mass stars with masses determined from orbital dynamics. Comparing
them to masses derived from different evolutionary models, they found
the masses of low-mass pre-main sequence stars to be generally
under-predicted. The mass at which model predictions are close to 
dynamically derived masses depends on the model, but agreement tends to 
be better above 0.5 M$_\odot$. The lowest tested pre-main sequence mass,
however, was 0.3 M$_\odot$. At main-sequence ages, \cite{Hillenbrand04} 
found masses to be under-predicted in the 0.1-0.6 M$_\odot$ regime in 
most of the models they used.  

The first study focusing on less massive objects at very young ages
was done by \cite{Mohanty04a,Mohanty04}, who derived masses for 11 low
mass objects in the young Upper Scorpius association. These mass
estimates are independent of evolutionary models, relying instead on
comparison of high resolution spectra with synthetic model spectra to
obtain effective temperatures and surface gravities, then combining
these with measured luminosities and the known distance and age of the
star-forming region. They conclude that models under-predict the mass
of young low-mass objects with a given temperature and luminosity,
similar to the results of \cite{Hillenbrand04} at higher masses and
ages. More recently, a study using the more familiar orbital dynamical
technique also suggested that the models under-predict mass for the
object AB~Dor~C \citep{Close05}. In that case, the determined mass of
0.09 M$_\odot$ is almost twice that which models predict for an object
of that luminosity, temperature, and age. Neither the age nor
temperature are determined very precisely by this study, however (see
also \cite{Luhman05}). \cite{Close05} overextend their result to
claim that low mass brown dwarfs are being mistaken for
``free-floating planets'' due to the mass calibration problem.
Actually, \cite{Mohanty04} find that models \emph {over-predict}
masses below about 0.03 M$_\odot$ (a regime not otherwise tested); the
slope of their empirical mass-luminosity relation is shallower than
predicted by models throughout the brown dwarf range.

Astronomers are much more used to calibrating masses by dynamical
studies of binary systems than by surface gravity analysis.  Indeed,
the dynamical method looks much more straightforward at first glance,
although when used to calibrate models it suffers from the same need
to accurately measure a star's luminosity and temperature as does the
surface gravity method.  In this paper, we report the discovery of a
double-lined spectroscopic binary among the same population of
low-mass members of Upper Sco studied by \cite{Mohanty04a,Mohanty04}.
This system is substantially younger than AB~Dor~C \citep[studied
by][]{Close05}.  We are able to determine an orbit for this system,
UScoCTIO~5, which allows us to place a lower limit on its mass.  We
show that its mass appears to be underpredicted by the models, based
on its measured luminosity and known age of the Upper Scorpius
association. We find the same result using temperature, obtained in
the usual way (via spectral type), which is not as precise as we would
desire.

Our current result tends to confirm the results suggested by the
surface gravity analysis of \cite{Mohanty04}, which are in the same
sense as found by \cite{Hillenbrand04} and \cite{Close05}.  This
strengthens the case that evolutionary models for young very low-mass
objects under-predict their masses at the bottom of the main sequence
and in the heavier half of brown dwarfs.  This has further
implications for the initial mass functions derived in young clusters
and star-forming regions, and provides a direction for modelers to
improve their initial conditions.  That they might need to do so comes
as no surprise; they have been saying so themselves
\citep[e.g.][]{Baraffe03}.

\section{Observations}

\begin{figure}  
\plotone{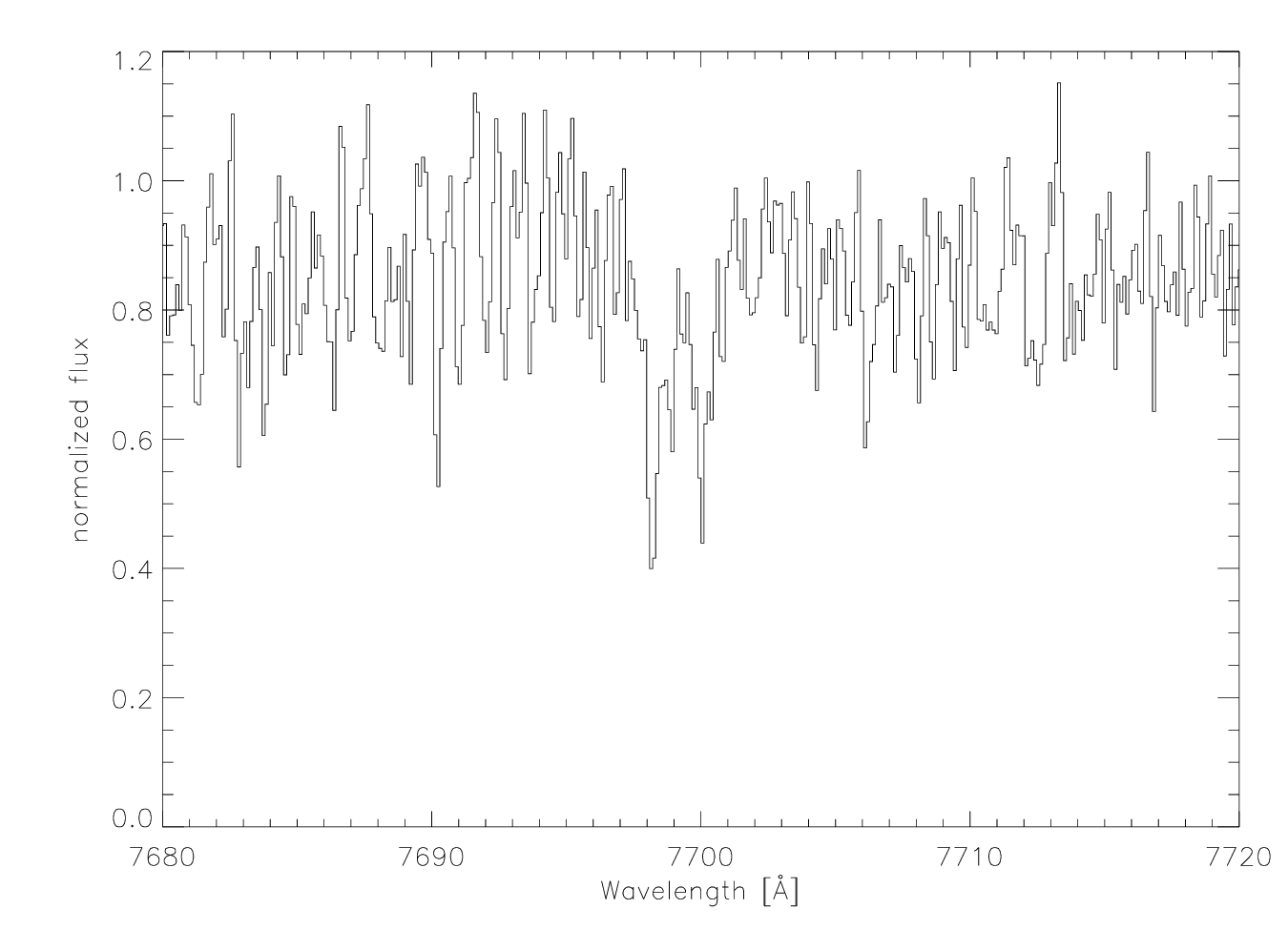}  
\caption{\label{fig:spectrum}Spectrum of UScoCTIO~5 taken with
  HIRES/Keck on June 11, 2003. The two components are clearly
  distinguishable in the K-line at 7700\,\AA.}
\end{figure}

UScoCTIO~5 was the warmest object we observed in an extension of our
original program to study very young brown dwarfs \citep{ Jayaward02,
  Jayaward03}.  We made the first observation of it on 11 June 2003
with HIRES at Keck~I. It was immediately clear from the raw spectrum
on that night that the lines were doubled. We then undertook a program
to determine the orbital parameters. The region around the KI lines in
the first spectrum is plotted in Fig.\,\ref{fig:spectrum}.  Both
binary components are clearly visible with comparable line  
intensities. After this exposure, twenty further spectra were taken in
2004, and one in 2005.  We (and other helpful observers) carried out
high-resolution spectroscopy using the HIRES spectrograph at Keck~I,
the echelle spectrograph at the CTIO~4m telescope, and the MIKE-Red
echelle on Magellan~II at Las~Campanas Observatory. The instruments
used, dates and times of observations, exposure times and names of
observers are compiled in Table\,\ref{tab:observations}.  For every
instrumental setup, one or more spectra of the slowly rotating M6-star
Gl~406 (CN~Leo) were also taken to allow production of
cross-correlation profiles.  The resolution in all our spectra is at
least $R = 25\,000$.

It is essential to the utility of our analysis that UScoCTIO~5 be a 
member of the Upper Scorpius Association. It is this membership 
which gives us the distance and age of the object. There are a number
of observations which strongly point to its membership. It was discovered
because it sits on the pre-main sequence for that association based on
its photometry (and of course is spatially within the association). 
If it is that young, it should display strong lithium lines, and indeed 
both components are easily visible in lithium. At 0.3 M$_\odot$, 
the lifetime of lithium is only around 30 Myr. This rules
out UScoCTIO~5 being a foreground main sequence star. There is further
evidence against that in the narrow Na line profiles (main sequence
stars have higher gravity and broader lines). The system velocity is
consistent with our other targets in Upper Sco; the central
velocity matches that of most other Upper Sco targets to within 3\,
km\,s$^{-1}$. Finally, the H$\alpha$~ lines in these objects are quite
similar to those in several of the other non-accretors in Upper Scorpius
shown in \cite{Mohanty04}, though that by itself would not be definitive.

\begin{deluxetable}{cccccr}
  \tablecaption{\label{tab:observations} Observation log}
  \tablewidth{0pt}
  \tablehead{\colhead{Instrument} & \colhead{Date} & \colhead{UT} & \colhead{exp. time [s]}& \colhead{$\Delta v_{\mathrm{rad}}$\,[km/s]}  & \colhead{Obs.} }
  \startdata
  HIRES/Keck  & 2003-06-11 & 10:15:08 &  900 & 70.2 & \tablenotemark{1}\\
  \tableline
  HIRES/Keck  & 2004-05-09 & 10:49:00 &  900 & 12.3 & \tablenotemark{2}\\
  HIRES/Keck  & 2004-05-10 & 11:45:14 & 1200 & 27.0 & \tablenotemark{2}\\
  HIRES/Keck  & 2004-05-11 & 09:18:10 & 1200 & 38.6 & \tablenotemark{1}\\
  HIRES/Keck  & 2004-05-11 & 11:48:19 &  900 & 41.4 & \tablenotemark{1}\\
  HIRES/Keck  & 2004-05-11 & 15:02:32 &  400 & 42.7 & \tablenotemark{1}\\
  \tableline
  echelle/CTIO 4m   & 2004-05-29 & 03:19:32 & 1800 & -37.7 & \tablenotemark{3}\\
  echelle/CTIO 4m   & 2004-05-29 & 03:57:40 & 1800 & -38.3 & \tablenotemark{3}\\
  echelle/CTIO 4m   & 2004-05-29 & 04:32:15 & 1800 & -38.8 & \tablenotemark{3}\\
  echelle/CTIO 4m   & 2004-05-30 & 04:04:57 & 1800 & -39.9 & \tablenotemark{3}\\
  echelle/CTIO 4m   & 2004-05-31 & 03:38:52 & 1800 & -42.9 & \tablenotemark{3}\\
  echelle/CTIO 4m   & 2004-05-31 & 04:10:55 & 1800 & -42.6 & \tablenotemark{3}\\
  echelle/CTIO 4m   & 2004-05-31 & 04:43:16 & 1800 & -42.3 & \tablenotemark{3}\\
  echelle/CTIO 4m   & 2004-05-31 & 08:06:56 & 1800 & -42.0 & \tablenotemark{3}\\
  echelle/CTIO 4m   & 2004-06-01 & 06:34:18 & 1800 & -42.8 & \tablenotemark{3}\\
  echelle/CTIO 4m   & 2004-06-03 & 01:23:55 & 1800 & -42.1 & \tablenotemark{3}\\
  echelle/CTIO 4m   & 2004-06-03 & 05:57:54 & 1800 & -41.1 & \tablenotemark{3}\\
  echelle/CTIO 4m   & 2004-06-03 & 06:30:39 & 1800 & -41.5 & \tablenotemark{3}\\
  echelle/CTIO 4m   & 2004-06-03 & 08:02:13 & 1800 & -41.4 & \tablenotemark{3}\\
  \tableline
  MIKE-Red/Magellan & 2004-06-13 & 03:59:21 & 1800 & 22.2 & \tablenotemark{4}\\
  MIKE-Red/Magellan & 2004-06-14 & 02:51:57 & 1800 & 35.0 & \tablenotemark{4}\\
  \tableline
  HIRES/Keck  & 2005-03-02 & 13:20:24 &  900 & -42.3 & \tablenotemark{5}\\
  \enddata
  \tablenotetext{1}{Basri}
  \tablenotetext{2}{Sargent, Takada-Hidai}
  \tablenotetext{3}{Mohanty, Tanner}
  \tablenotetext{4}{Paulson}
  \tablenotetext{5}{Basri, Reiners}
\end{deluxetable}

\section{Differential Radial Velocities}
\label{sec:analysis}

We infer the orbital elements of UScoCTIO~5 from the radial velocities 
of the binary components relative to each other (differential radial 
velocities). Cross-correlation functions are computed using spectra of 
Gl~406 as a template. For each spectrum, we obtained a template spectrum 
using identical instrument setups for UScoCTIO~5 and Gl~406.
In case of HIRES spectra, we used one order covering approximately
100\,\AA\ between 7850\,\AA\ and 7950\,\AA. From the CTIO data we
merged four echelle orders and calculated the correlation functions in
about 600\,\AA\ between 8350\,\AA\ and 8950\,\AA. The region between
8600\,\AA\ and 8800\,\AA\ was used in the Magellan data. None of these
wavelength regions is contaminated by significant telluric lines.

Using wavelength regions as large as 600\,\AA, the wavelength
dependency of Doppler shifts becomes important for the calculation of
precise radial velocities. At differential radial velocities of
50\,km\,s$^{-1}$, for example, the velocity difference per pixel is of
the order of 3\,km\,s$^{-1}$ from one end of the linear wavelength
scale used in the CTIO data to the other end. For highest accuracy, we
therefore calculate the correlation function using logarithmic
wavelengths.  Differential radial velocities are then derived from the
separation of the two peaks belonging to the different components.
Both components were easily distinguishable in all correlation
functions. Differential radial velocities $\Delta v_{\mathrm{rad}}$
for all observations are given in column five of
Table\,\ref{tab:observations}. An estimation of our measurement errors
will be given in the next section.

It turns out that all observations taken after our first one in 
2003 reveal significantly lower differential radial velocities, i.e.,
the components are more seriously blended than in the example given in
Fig.\,\ref{fig:spectrum}. While both components are always
distinguishable, an assignment of each of the two correlation peaks to
individual stellar components is ambiguous, even in the
cross-correlation functions derived from large spectral regions. We
therefore did not individually identify the components in every
spectrum, i.e., the sign of $\Delta v_{\mathrm{rad}}$ is not derived
from the data. However, as shown in the next section, we were able to
eliminate the ambiguity in the sign of $\Delta v_{\mathrm{rad}}$ when
determining the orbit of UScoCTIO~5.

\section{Determining the Orbit}

As mentioned above, we do not have information about the sign of
$\Delta v_{\mathrm{rad}}$. Our data were taken during five observing
campaigns, indicated by horizontal lines in
Table\,\ref{tab:observations}. Radial velocities from spectra obtained
during the same campaigns show clear correlations, and it is very
unlikely that the sign of $\Delta v_{\mathrm{rad}}$ changes within one
campaign.  However, we looked at the periodograms of all possible
permutations of the signs of $\Delta v_{\mathrm{rad}}$ and found no
hints of a short orbit with the sign of $\Delta v_{\mathrm{rad}}$
alternating within consecutive observations of any campaign. The
radial velocity curve of a spectroscopic binary can be uniquely
specified by the following five parameters: Period, $P$; total
projected mass $(M_1 + M_2)\sin{i}$; eccentricity of the orbit, $e$;
longitude of periastron, $\Omega$; and Epoch, $T$. From differential
radial velocities it is not possible to obtain the inclination $i$.
Thus we only derive a minimum mass sum from our data.  The mass
fraction $M_1/M_2$ could be determined in principle from either the
individual component line ratios or the amplitude of their excursions
about the system velocity. In our case, we could not confidently find
any difference between the two components. This places a minimum mass
limit on the primary (it cannot contain much more than half of the
total mass).

\begin{figure}  
  \epsscale{1.0} \plottwo{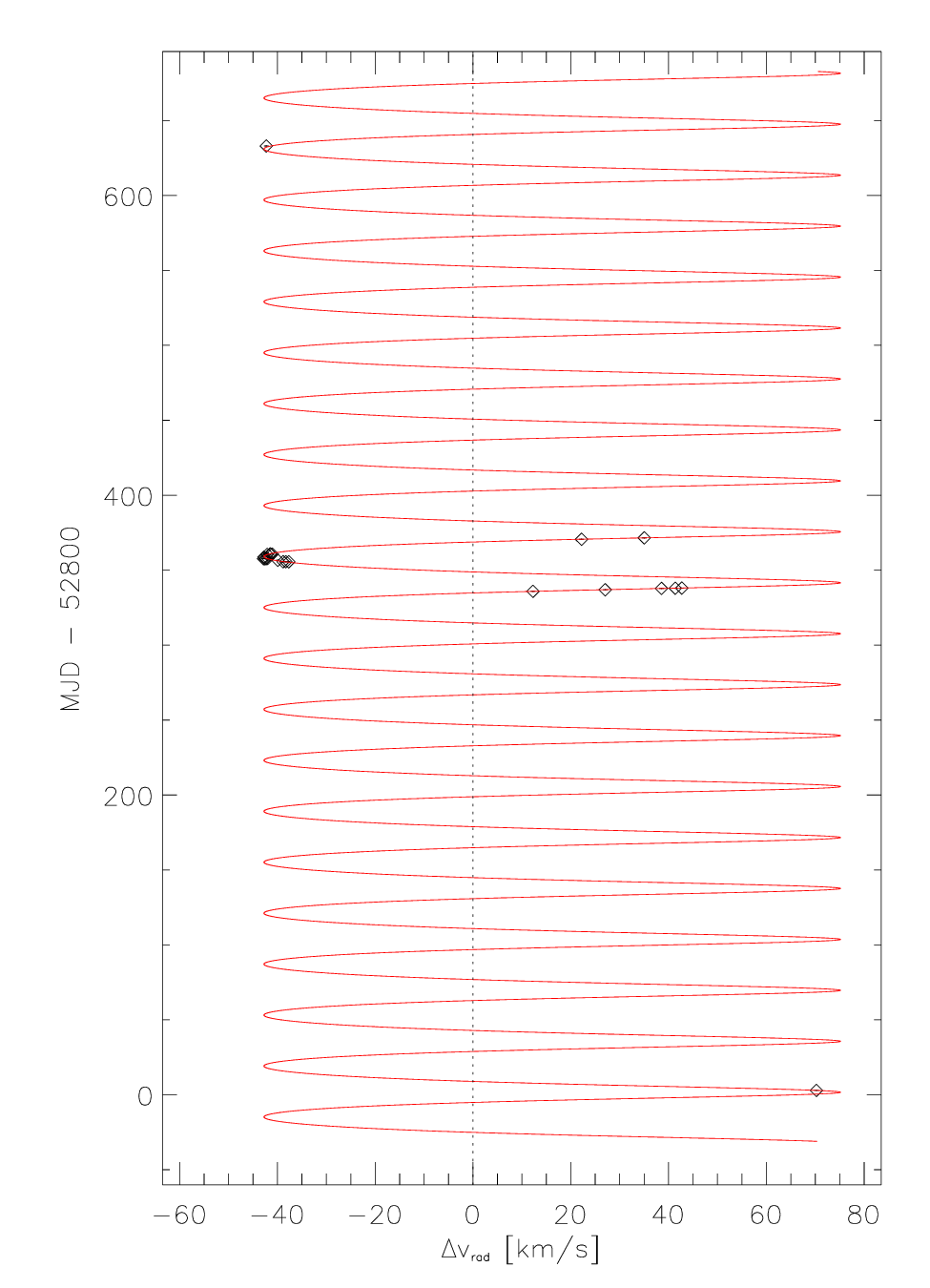}{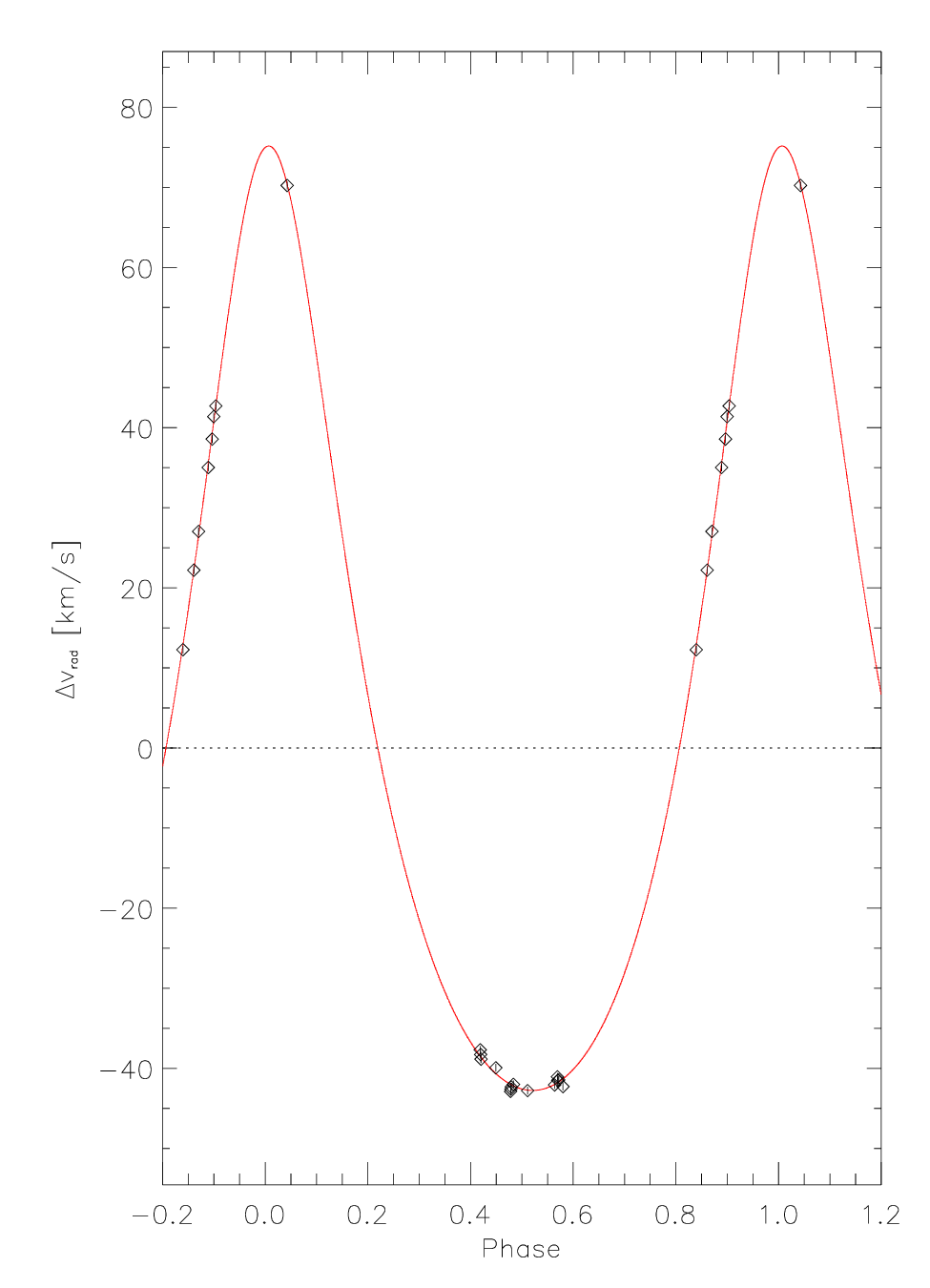}
  \caption{\label{fig:orbit}Measurements of differential radial
    velocity $\Delta v_{\mathrm{rad}}$ (symbols) and orbital solution
    (red line).  \emph{Left panel}: $\Delta v_{\mathrm{rad}}$ vs. MJD$
    - 52\,800$.  \emph{Right panel}: Phase binned $\Delta
    v_{\mathrm{rad}}$.  The rms scatter of $\Delta v_{\mathrm{rad}}$
    is 480\,m\,s$^{-1}$; these errors are smaller than the symbol
    size.}
\end{figure}

We searched for the best orbital solution in the five free parameters
by minimizing the rms scatter of $\Delta v_{\mathrm{rad}}$, $\sigma^2
= \sum (\Delta v_{\rm rad} - \Delta v_{\rm rad, orbit})^2/N$, with $N
= 22$ the number of measurements. The rms scatter of $\Delta
v_{\mathrm{rad}}$ around the best fit is $\sigma_{\rm min} =
480$\,m\,s$^{-1}$. The uncertainties in $\Delta v_{\mathrm{rad}}$ are
not well determined due to the variety of instruments used, and they
are primarily systematic rather than of statistical origin. Thus we 
cannot give a strict statistical confidence limit on our result. 
We conservatively estimate the uncertainty in the orbital parameters 
by searching for the intervals for which $\sigma < 2\sigma_{\rm min}$, 
when varying each parameter independently.  The parameters of our best 
fit are given in Table\,\ref{tab:orbit} with the derived errors. 
Our best solution is plotted 
over the whole range of observations and over phase in the left and 
right panels of Fig.\,\ref{fig:orbit}, respectively.

We find a minimum projected total mass of $(M_1 + M_2)\sin{i}
=0.64$\,M$_\odot$.  The solution is unique in the sense that for the
second best local minimum in $\sigma$ we get a significantly higher
value of $\sigma_2 = 7.5\sigma_\mathrm{min}$. This solution would
provide $(M_1 +M_2)\sin{i} = 0.70$\,M$_\odot$, and all other
parameters are comparable to our best solution as well. We ran an
extensive survey in parameters considering different signs in
individual values and groups of $\Delta v_{\mathrm{rad}}$. No other
parameter set yielding comparable fit quality could be found.

\begin{deluxetable}{lrcl}
  \tablecaption{\label{tab:orbit} Orbital solution; errors denote
    values at which $\sigma^{2} = 2\sigma^{2}_\mathrm{min}$ }
  \tablewidth{0pt} \tablecolumns{4} \tablehead{\colhead{Parameter} &
    \multicolumn{3}{c}{Value}} \startdata
  Period, $P$ [d] & $33.992$ & $\pm$ & $0.006$\\
  $(M_1 + M_2)\,\sin{i}$ [M$_\odot$] & $0.64$ & $\pm$ & $0.02$\\
  Eccentricity, $e$ & $0.276$ & $\pm$ & $0.008$ \\
  Semimajor axis, a [AU]& $0.177$ & $\pm$ & $0.002$\\
  Longitude of periastron, $\Omega$ [$\degr$] & $274.5$ & $\pm$ & $0.8$\\
  Epoch (MJD), $T$ [d] & 52\,799.974 & $\pm$ & 0.002\\[2mm]
  \enddata
\end{deluxetable}

\section{Comparison to Evolutionary Tracks}

In order to compare the derived minimum mass to theoretical
evolutionary tracks, we calculate bolometric luminosity
$L_{\mathrm{bol}}/L_{\odot}$ and effective temperature
$T_{\mathrm{eff}}$ from photometry and spectral type, respectively.  A
spectral type of M4 has been determined for UScoCTIO~5 by
\cite{Ardila00}.  Photometry, effective temperatures and bolometric
luminosity are given in Table\,\ref{tab:upsco5}.  We calculate
bolometric luminosity from $J$ and $K$ colors taken from 2MASS
\citep{2MASS}, and determine $L_{\mathrm{bol}}/L_{\odot}$
independently from both colors. From the $J$-color we determine $V$
according to dwarf calibrations given in \cite{Kenyon95}, $BC_V$ is
also taken from that paper. A calibration of $BC_K$ in the UKIRT
photometric system can be found in \cite{Leggett01}; the
transformation of 2MASS colors to the UKIRT system is adopted from
\cite{Carpenter01}.  

We calculate the extinction by comparing $J-H$
and $J-K$ to the young-disk calibration given in \cite{Leggett92}
using the extinction law from \cite{Schlegel98}; colors are converted
to the $CRI$-system for that purpose \citep{Carpenter01}. This yields
$A_V(J-H) = 1.0$ and $A_V(J-K) = 0.5$, while a value of $A_V = 0.5 \pm
0.5$ is reported in \cite{Ardila00}. Including the uncertainty in
spectral type, our estimates of $A_V$ have uncertainties of 0.3\,mag,
hence the values of $A_V$ are consistent within the uncertainties.  We
choose $A_V =0.75 \pm 0.4$ for the calculation of bolometric
luminosity, and we derive a bolometric luminosity of
$\log\,L_{\mathrm{bol}}/L_{\odot} = -1.17 \pm 0.08$ from the $J-$band
and $\log\,L_{\mathrm{bol}}/L_{\odot} = -1.16 \pm 0.06$ from the
$K-$band. The two values derived from different colors agree very
well, and in the following we will use the luminosity derived from the
$K$-band since it has smaller uncertainties. For the uncertainties, we
took into account a spectral class uncertainty of half a subclass
($\pm0.5$), age ($5\pm 2$ Myr) and distance ($145\pm 20$ pc) for the
cluster adopted from \cite{Mohanty04}. These errors have been
included in the luminosity error. For the bolometric correction $BC_V$
an uncertainty of 0.1\,mag has been assumed, and 0.05\,mag has been
employed for $BC_K$ \citep{Leggett01}.

\begin{figure}  
  \epsscale{1.0} \plotone{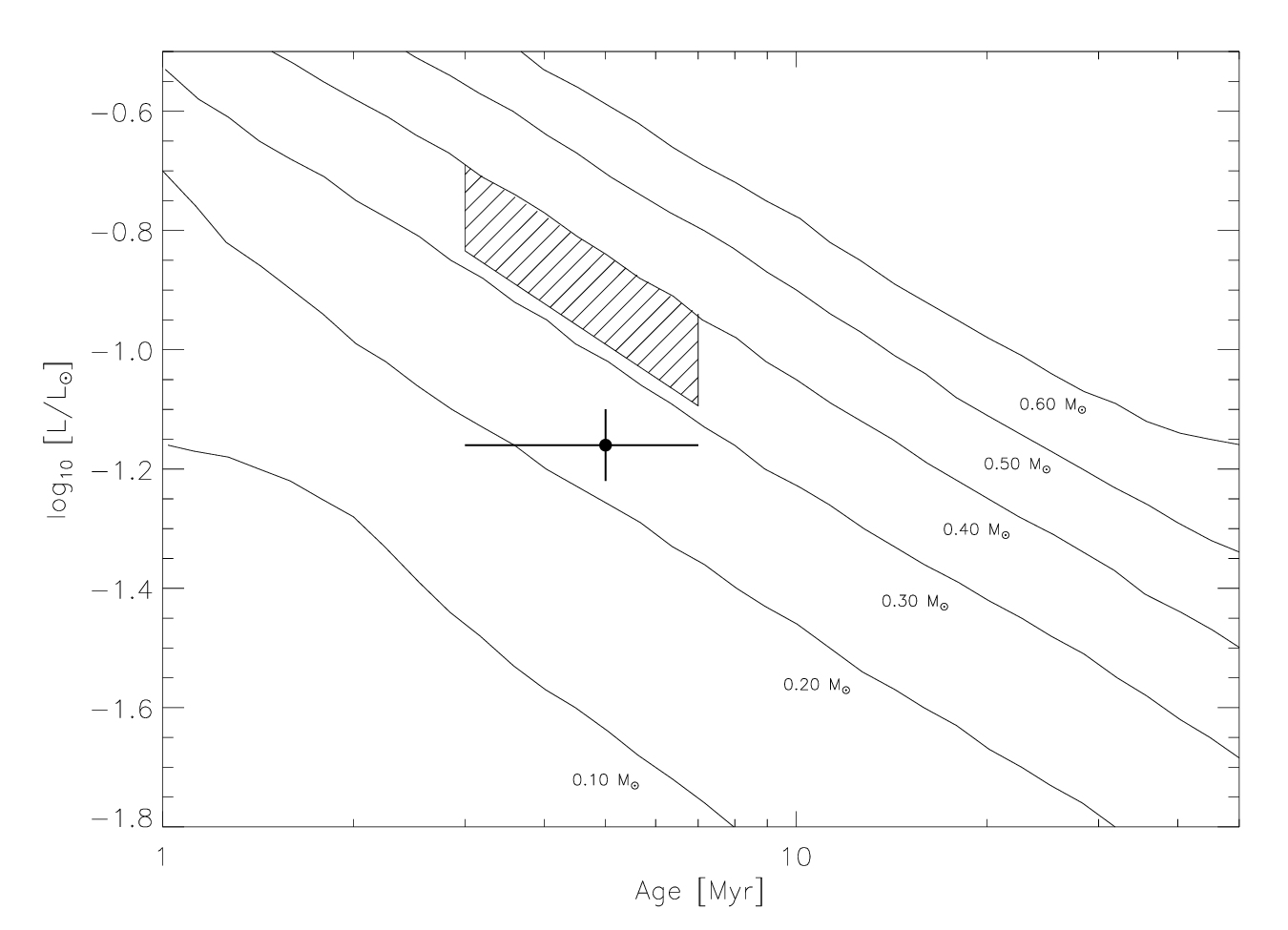}
 \caption{\label{fig:LumAge}
 A luminosity-age diagram with evolutionary tracks for various masses 
(solid lines) from \cite{Baraffe98} and \cite{Chabrier00}.  The
 point shows the luminosity of the primary of UScoCTIO~5 (and its
 uncertainty), derived from its J-band luminosity and the distance to the
 association.  Age estimates for the association define the uncertainties
 in the abscissa.  The hatched region shows the allowed mass of UScoCTIO~5
 set by its dynamical lower limit and the upper limit due to mass-spectral
 type relations on the main sequence.  There is a clear underestimation of
 its mass (or overestimation of its luminosity, given the true mass) by
 the models.  }
\end{figure}

In Fig.\,\ref{fig:LumAge}, UScoCTIO~5 is plotted in a luminosity-age
diagram with evolutionary tracks from \cite{Baraffe98} and
\cite{Chabrier00}.  Assuming that both components of UScoCTIO~5 contribute
equal light, the most probable mass of one component according to its
position in the diagram is $M = 0.23$\,M$_{\odot}$.  Our dynamically
determined total system projected minimum mass is $(M_1+M_2)\sin{i} = 0.64
\pm 0.02$\,M$_{\odot}$.  The color and spectral type of a spectroscopic
binary is primarily determined by the hotter component, i.e., the minimum
mass of the more massive component in the binary system is at least
\begin{equation}  
  M_{\mathrm{min}} = 0.5(\mathrm{M}_1+\mathrm{M}_2) = \frac{0.32}{\sin{i}}\,\mathrm{M}_{\odot}. 
\end{equation}
The real mass is very likely to be higher because of the inclination
effect. On the other hand, the inclination is probably fairly close to
edge-on, because the lower limit on mass is already rather high for the spectral type (this luck allows us to find a meaningful lower limit). 
The hatched region in Fig.\,\ref{fig:LumAge} is defined by the 
known age of the association and the lower limit to the dynamical mass 
from the orbit. The upper limit to the mass is set from the late 
spectral type.  Even on the main sequence, an M2 dwarf has a mass 
of about $M = 0.4$\,M$_{\odot}$  \citep{Delfosse00}, the upper limit 
we adopt.  Our object is both cooler and younger than an M2 dwarf, 
making our upper limit a safely conservative estimate. 

Temperatures of young dwarfs have been investigated by \cite{Luhman99} and 
\cite{Luhman03}.  Investigating members of the young cluster IC~348 and
the young quadruple system GG~Tau, they provide a temperature scale
calibrated by evolutionary models under the premises that coeval stars
fall on the same isochrones.  These temperatures are intermediate between
older giant and dwarf temperature scales, leading to an estimate for the
temperature(s) of UScoCTIO~5 of $T_{\mathrm{eff}} \approx 3270\pm100$\,K.
The error here is that cited by Luhman, and is his estimate of possible  
systematic errors in his scale (which was constructed to fill a 
specific purpose, relevant in our context).
\cite{Mohanty04a} have discussed the temperatures of the cool components of
GG~Tau in detail (their Appendix B). They have the advantage of a
spectroscopically determined temperature in addition to newer spectral types.
They find that the temperatures of these young objects (comparable to
UScoCTIO~5) are consistent with newly determined dwarf scales for main
sequence M-dwarfs given in \cite{Leggett00} or \cite{Mohanty03}.  This
leads to an estimate for the temperature of UScoCTIO~5 (M4) of
$T_{\mathrm{eff}}\approx 3175\pm100$\,K. 

\begin{figure}  
  \epsscale{1.0} \plotone{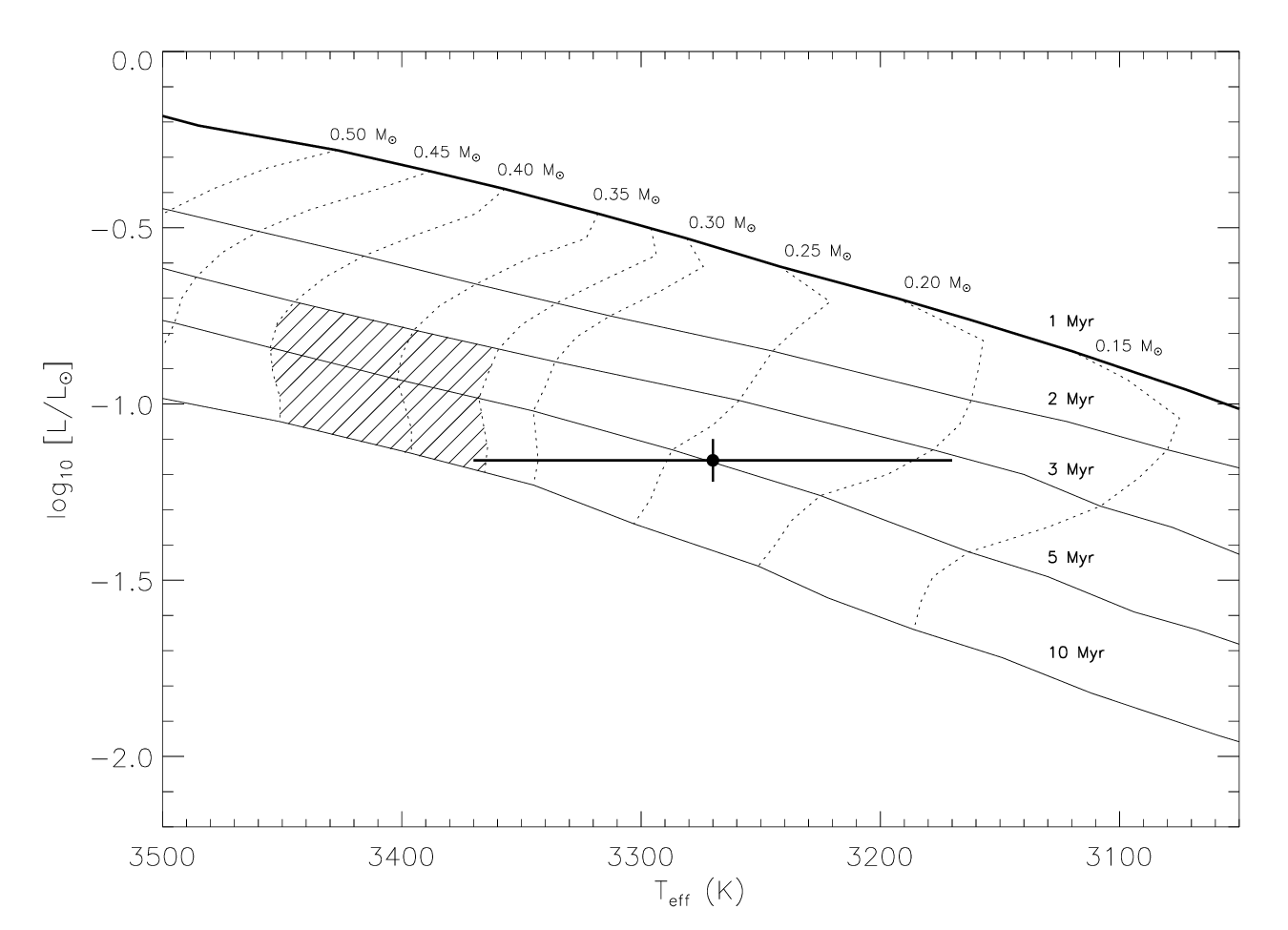}
 \caption{\label{fig:HRD}
 An HR-diagram with isochrones (solid lines) and evolutionary tracks
 (dotted lines) from \cite{Baraffe98} and \cite{Chabrier00}.  The point
 shows the temperature of UScoCTIO~5 derived from its spectral type and
 the temperature scale of \cite{Luhman03}; we adopt their uncertainties 
 of $T_{\mathrm{eff}} (\pm 100$\,K$)$.  The hatched region shows
 what is allowed by the age limits on the association and the mass limits
 from the orbit of UScoCTIO~5 and its late spectral type.  
 The result here is very similar to that from the luminosity-age diagram.}
\end{figure}

In Fig.\,\ref{fig:HRD}, UScoCTIO~5 is plotted in an HR-diagram with
evolutionary tracks from the same models.  With the effective temperature
derived from a scale that is designed to fit isochrones of these
evolutionary models, it is not surprising that UScoCTIO~5 is located near
the 5\,Myr isochrone.  If we adopt the temperature scale based on the work
of \cite{Mohanty03} and \cite{Mohanty04a} which is about 100K cooler, the
effect is to lower the inferred mass to about $M = 0.19$\,M$_{\odot}$ and
the age to 4\,Myr (still fully consistent with the age estimated for the
association).  An equally weak link in setting the temperature is 
the assumed spectral type.  A change to M3.5 from M4, for example, 
would have a substantial effect on the implied temperature, and raise the 
inferred mass to better agreement with the models.  We intend to determine 
the temperature of UScoCTIO~5 directly from the high resolution spectra in 
a follow-up paper. It is therefore the case that the mass discrepancy 
cannot be demonstrated as conclusively in the HR-diagram with the current
uncertainties in temperature, and our conclusions rest primarily on the
(similar) result in the luminosity-age diagram.

\begin{deluxetable}{lrcl}
  \tablecaption{\label{tab:upsco5} UpSco~5}
  \tablewidth{0pt}
 \tablehead{\colhead{Parameter} & \multicolumn{3}{c}{Value}}
  \startdata
  R.A. (J2000.0)  & \multicolumn{3}{c}{\phs$15\,59\,50.39$}\\
  Decl. (J2000.0) & \multicolumn{3}{c}{$-19\,44\,35.8$\phn}\\
  Spectral Type\tablenotemark{a} & M4 & $\pm$ & $0.5$\\
  Distance \tablenotemark{a}  &$145$\,pc & $\pm$ & $20$\\
  $J$\tablenotemark{b}   & $11.17$ & $\pm$ & $0.02$\\
  $H$\tablenotemark{b}   & $10.45$ & $\pm$ & $0.03$\\
  $K$\tablenotemark{b}   & $10.17$ & $\pm$ & $0.02$\\
  $A_V$\tablenotemark{a} & $0.75$ & $\pm$ & $0.4$\\
  \tableline
  log\,$L_{\mathrm{bol, J}}/L_{\odot}$ & $-1.17$ & $\pm$ & $0.08$\\
  log\,$L_{\mathrm{bol, K}}/L_{\odot}$ & $-1.16$ & $\pm$ & $0.06$\\
  $T_{\mathrm{eff}}$\tablenotemark{a,d}\quad [K]& $3270$ & $\pm$ & $100$\\ 
  Age & 5 Myr &$\pm$ & 2 Myr \\
  \enddata
  \tablenotetext{a}{\cite{Ardila00}}
  \tablenotetext{b}{\cite{2MASS}}
  \tablenotetext{c}{see text}
  \tablenotetext{d}{\cite{Luhman03}} 
  
  \tablecomments{$T_{\mathrm{eff}}$ was derived from a scale
    intermediate between dwarf and giant temperatures, using a dwarf
    scale yields $T_{\mathrm{eff}} = 3180$\,K \citep{Luhman99}.}
\end{deluxetable}

The hatched region in Fig.\,\ref{fig:HRD} displays the area consistent
with our previously adopted mass and age limits. The position of UScoCTIO~5 
inferred for its dynamically determined mass from evolutionary models is 
too hot to be consistent with its measured luminosity. With the error 
bars on temperature (assuming a spectral type of M4) adopted from 
\cite{Luhman99}, UScoCTIO~5's position in the HRD is barely consistent with 
even the minimum mass at $i = 90^\circ$ and equal mass components. 
The conclusion from Fig.\,\ref{fig:HRD} is that the mass of UScoCTIO~5 is 
probably significantly underestimated by the evolutionary model parameters 
that would fit the observational temperature and luminosity.  
The mass estimated by models in the luminosity-temperature diagram is 
consistent with that estimated from the luminosity-age diagram, because the 
temperature scale employed positions objects near the 5 Myr isochrone by 
construction. This probably means that the temperature is not too far off.

Besides a problem with the evolutionary models themselves, the 
underestimation of mass could only be due to (1) uncertainty in the 
dynamical mass, (2) underestimation of the temperature, and/or (3) 
mis-estimation of the spectral type by a half-subclass cooler or more. The 
dynamical mass does not have a large error, except that the unknown 
inclination can easily make the mass discrepancy worse.  The estimation of 
luminosity is unlikely to have a large error, and this only plays 
a minor role since evolutionary tracks are largely independent of luminosity at a given mass; they are nearly vertical in the hatched region
in Fig.\,\ref{fig:HRD}.  An error in the spectral type, or in the
conversion of spectral type to temperature, could move the observed point
leftward in Fig.\,\ref{fig:HRD}.  Note, however, that moving it to the
value of 3450\,K which is suggested by the dynamical mass, would also
make the age uncomfortably large (as the luminosity would be too low).
We leave the detailed analysis of these issues to another paper, wherein
we will obtain an independent spectroscopic determination of the temperature. 
Our current result from the luminosity-age analysis is not affected by 
errors in spectral type or its conversion to temperature in any case.

\section{Conclusions}

We have derived the orbit of the M4 binary system UScoCTIO~5, a member of
the Up\,Sco OB association with an age of $\sim5$\,Myr.  From our radial
velocity measurements we infer a projected minimum total mass of the
system of $(M_1 + M_2)\,\sin{i} = (0.64 \pm 0.02)$\,M$_\odot$, i.e.  a
minimum mass of the primary of $M = 0.32$\,M$_{\odot}$. The luminosity and
effective temperature of UScoCTIO~5 have been derived from photometry and
spectral type.  As our main result, we compare the position of UScoCTIO~5
to theoretical evolutionary tracks of \cite{Baraffe98} and
\cite{Chabrier00} in a luminosity-age diagram.  These evolutionary tracks
predict a mass of $M \approx 0.23$\,M$_{\odot}$ at the luminosity and age
observed for UScoCTIO~5.  Taking into account our ignorance of the
inclination $i$ of the orbit, and the mass ratio $M_2/M_1$ (both of which
can only lead to higher dynamical mass), we conclude that there is a
significant discrepancy between the ``true'' mass and the mass
derived from current commonly-used evolutionary models in this part of
parameter space.

This conclusion is the same in our analysis using the HR-diagram,
although there it is subject to further uncertainties in the true
spectral type, and the conversion from spectral type to temperature at
this age.  Uncertainties in the derivation of $T_\mathrm{eff}$ lead to
lower masses and an even larger discrepancy with the evolutionary
models if the temperature is cooler than we estimate here; there is
some suggestion for that from \cite{Mohanty04a}.  One could obtain
higher mass (and better agreement with the models) if the spectral
type were earlier than the current value by half a subclass or more
(using the current temperature scale).  The same issues are present in
the analysis by \cite{Close05}, who also found that models
underestimated the mass for an object that is lower in mass but
substantially older than this one. These uncertainties, however, do
not affect the conclusions of the first paragraph.

These problems with temperatures are not applicable to the
similar conclusions about model discrepancies reached earlier in the
analysis of a number of objects in Upper Sco having even lower masses by
\cite{Mohanty04}, because they determined the temperatures of those
objects more directly (from high resolution spectra).  However, their
masses are found by a surface gravity analysis rather than dynamically.
The effect of the current analysis, therefore, is to support the
methodologies of the earlier paper with an independent check. It seems fair to
say at this point that the evidence is accumulating from several
directions that commonly-used evolutionary models require adjustments to
higher mass at a given luminosity and/or temperature for stars or brown
dwarfs in the 0.03-0.3 solar mass range which are younger than a 
hundred million years or so.  This has implications for all the conclusions
drawn about the initial mass function in star-forming regions at low
masses, and for other papers which rely on evolutionary mass estimates to
draw conclusions about what sort of low-mass objects are being studied.

\acknowledgments 
We would like to acknowledge the kind assistance of
several observers who willingly aided this project during time they
had for other purposes. They include Wallace Sargent and Masahide
Takada-Hidai from CIT, Angelle Tanner from SAO, and Diane Paulsen
from U~Mich. This work is based on observations obtained from the W.M. Keck
Observatory, which is operated as a scientific partnership among the
California Institute of Technology, the University of California and
the National Aeronautics and Space Administration, on data taken at
Las Campanas Observatory, and at CTIO which is operated by AURA for the NSF. 
We would like to acknowledge
the great cultural significance of Mauna Kea for native Hawaiians and
express our gratitude for permission to observe from atop this
mountain. GB thanks the NSF for grant support through AST00-98468. AR
has received research funding from the European Commission's Sixth
Framework Programme as an Outgoing International Fellow (MOIF-CT-2004-002544).


\end{document}